\documentclass{article}
\input{epsf.sty}

\def\abstract#1{\vskip 7mm 
        \begin{center}{\large Abstract}\par \smallskip
                \begin{minipage}[c]{12cm}
                        \small #1
                \end{minipage}
        \end{center}
}
\def\title#1{\begin{center}{\Large\bf #1}\end{center}}
\def\author#1{\vskip 5mm \begin{center}{#1}\end{center}}
\def\address#1{\begin{center}{\it #1}\end{center}}
\makeatletter
\@ifundefined{lesssim}{}{}
\@ifundefined{gtrsim}{}{}
\def\vereq#1#2{\lower3pt\vbox{\baselineskip1.5pt \lineskip1.5pt
\ialign{$\m@th#1\hfill##\hfil$\crcr#2\crcr\sim\crcr}}}
\makeatother

\begin{document}
\baselineskip12.0pt
\title{%
Application of the Separate Universe Approach to Preheating
}
\author{%
  Takahiro Tanaka${}^a$\footnote{E-mail:tanaka@yukawa.kyoto-u.ac.jp, 
 \hspace{1cm}$^{2}$E-mail:Bruce.Bassett@port.ac.uk \newline
 \indent
 This work was supported in part by Monbukagakusho Grant-in-Aid Nos. 14740165 and 14047214.}
 and 
  Bruce Bassett${}^{b2}$
}
\address{%
  $^a$Yukawa Institute for Theoretical Physics, 
  Kyoto University, Kyoto 606-8502, Japan\\
  $^b$Institute of Cosmology and Gravitation (ICG), 
  University of Portsmouth, \\
  Mercantile House, Portsmouth, PO1 2EG, England
}
\abstract{
  The dynamics of preheating after inflation has not been clearly understood 
yet.In particular, the issue of the 
generation of metric perturbations during preheating on 
super-horizon scale is still unsettled. Large scale perturbations 
may leave an imprint on the cosmic microwave background, or may become 
seeds for generation of primordial black holes. Hence, in order to 
make a connection between the particle physics models and cosmological 
observations, understanding the evolution of 
super-Hubble scale perturbations during preheating is important. 
Here, we propose an alternative treatment to handle this issue based on  
the so-called separate universe approach, 
which suggests less efficient amplification of
super-Hubble modes during preheating than was
expected before. We also point out an important issue which
may have been overlooked in previous treatments. 
}

\section{Introduction}
Rather recently, people realized that the process of reheating 
after inflation need not simply be described by the usual single 
particle decay of the inflaton field. 
If there are relevant scalar field degrees of freedom 
coupled to the inflaton field at the end of inflation, the 
coherent oscillation of the inflaton can excite those 
scalar fields by parametric resonance\cite{brandenberger}. 
This process was extensively investigated, in mast cases, neglecting
metric perturbations, which typically will be a good
approximation as long as we are interested in 
processes occurring within the Hubble horizon scale.  
Often the energy transfer due to parametric resonance is 
much more efficient than that due to the usual particle decay, and 
sometimes it is violent. It was then suspected that there 
might be significant effects on the large scale perturbations  
beyond the Hubble scale\cite{super-hubble} due to resonance. 
Some particle physics models may predict 
distinguishable signals in the cosmic microwave background, 
or formation of inconsistently large number of primordial black
holes \cite{TsuBas}. 

Here in this short contribution, we would like to propose another 
approach to discuss this issue of the generation of large 
scale perturbations at preheating. We make use of 
the separate universe approach~\cite{Wands}. We stress that unphysical 
acausal energy transfer, which may not carefully prohibited 
in previous treatments, does not take place in this 
new approach.

\section{Separate universe approach}
\underline{\it\large Basic Idea}$~$:$~$
We consider the system composed of many scalar fields. 
The discussions in this paper is based on the 
following statement obtained in Ref.~\cite{SasTan,Wands,KHNT}. 
\vspace*{3mm}
\newline
\hspace*{5mm}
\begin{minipage}{15cm}
\vbox{\baselineskip=10pt
When we solve the background spatially homogeneous 
field equation, the 
$e$-folding number $N\equiv \log a$ until the expansion rate $H$ reaches 
a given value $H_r$ depends on the 
initial value of the fields ${\mbox{\boldmath$\Phi$}}$ 
(and their time derivatives). 
Let's denote this value of $N$ by $N_{H_r}$. Suppose that the initial 
value of ${\mbox{\boldmath$\Phi$}}$ has spatial variation over 
super-Hubble scales. Then, the value of $N_{H_r}$ becomes 
a function of the spatial position ${\bf x}$. 
In this case, the newly generated 
curvature perturbations on the constant Hubble slice ${\cal R}$ 
is simply given by $\Delta N_{H_r}$, i.e., the fluctuations in $N_{H_r}$.  
}\end{minipage}

\vspace{3mm}
\newpage
\noindent
\underline{\it\large Mechanism of Generating Curvature Perturbations 
from Isocurvature Perturbations}$~$:$~$

\vspace{1mm}
Let's consider matter fields consisting of two components. 
We call them $\phi$ and $\chi$. They are not necessarily 
scalar fields here. We assume that the rates of change of the 
energy density of $\phi$ and $\chi$ due to the expansion 
of the universe are different. 
Namely,
\begin{equation}
 {d\log\rho_\phi\over dN} = -3(1+w_\phi),  \quad
 {d\log\rho_\chi\over dN} = -3(1+w_\chi),  
\end{equation}
For example, the energy density of $\phi$ 
changes like dust matter, while $\chi$ behaves 
as a radiation fluid. 
In this case, we have $w_\phi=0$ and $w_\chi=1/3$. 
Let us denote the fraction of the energy density of $\chi$ by 
$f:=\rho_\chi/\rho$. 
Then, we have 
\begin{equation}
 {d\log\rho\over dN} = -3(1+w_\phi)+3(w_\phi-w_\chi)f.   
\label{generalformula}
\end{equation}
From this expression, we notice that there are two factors necessary 
for generation of super-Hubble adiabatic perturbations. 
One is non-vanishing $w_\phi-w_\chi$, and the other is 
inhomogeneity of $f$.

\section{Conformal two field model}
\noindent
\underline{\it\large Model}$~$:$~$
We consider the model whose potential is given by 
\begin{equation}
 V={\lambda\over 4}\phi^4+{1\over 2}g^2\phi^2\chi^2. 
\label{conf-model}
\end{equation}
When $g^2=2\lambda$, the longest wavelength mode of $\chi$ is
the fastest growing mode,  
and the initial spectrum of $\chi$ is scale-invariant. This 
appears to lead to exponential growth of pre-existing entropy 
perturbations even
when backreaction is taken into account~\cite{BV}.  
The equations of motion for the spatially homogeneous background 
are given by
\begin{eqnarray}
\ddot \Phi^i+3H \dot\Phi^i
    -{dV\over d\Phi^i}=0,\qquad
H^2={\kappa^2\over 3}\left({1\over 2}
   \dot {\mbox{\boldmath$\Phi$}}^2+V\right).
\end{eqnarray}
Here we used the vector notation to represent the two 
component scalar fields, $\phi$ and $\chi$, simultaneously.  
Namely, $\Phi^1=\phi$ and $\Phi^2=\chi$. \newline

\noindent
\underline{\it\large {Simplified Analysis with Time Averaging}}$~$:$~$
We define the averaged energy in a unit comoving volume by 
$
 {\cal E}=\left\langle a^3
   \left({1\over 2}\dot {\mbox{\boldmath$\Phi$}}^2+V\right)\right\rangle,  
$
where $\langle\cdots \rangle$ represents time averaging. 
Let us assume that ${\mbox{\boldmath$\Phi$}}^2$ is bounded and is
oscillating. Then, we have many zeros of 
$\dot {\mbox{\boldmath$\Phi$}}\!\cdot\! {\mbox{\boldmath$\Phi$}}$. 
If we take the average between two zeros, we have
\begin{equation}
0=\left\langle{d\over dN}\left(a^3 \dot {\mbox{\boldmath$\Phi$}}
\!\cdot\! {\mbox{\boldmath$\Phi$}}
\right)\right\rangle=
 \left\langle {a^3\over H}
     \left(\dot {\mbox{\boldmath$\Phi$}}^2-\lambda\phi^4-2g^2\chi^2\phi^2
\right)\right\rangle. 
\end{equation}
Since $H$ is slowly changing, we obtain an approximate
relation: 
$
 \left\langle a^3 
     \left(\dot {\mbox{\boldmath$\Phi$}}^2-\lambda\phi^4-2g^2\chi^2\phi^2
\right)\right\rangle=0. 
$
Therefore we have 
\begin{equation}
 {d\log{\cal E}\over dN}=3-3{\left\langle a^3\dot {\mbox{\boldmath$\Phi$}}^2
  \right\rangle\over {\cal E}} \approx -1. 
\end{equation}
From this fact, ${\cal E}\approx e^{-N}$ is deduced, and hence we have
$
H^2\propto \rho
   \approx {\cal E}/a^3\propto a^{-4}.
$
Anyway, $N_{H_r}$
is basically independent of the initial value of the field. 
Therefore it seems that additional curvature perturbations are not 
generated when the wavelength of a mode is longer than the horizon scale. 
In the language of the preceding section, 
$w_\phi-w_\chi$ vanishes approximately in this model. 
Hence, amplification of
fluctuations of $O(1)$ is hardly expected. 
Notice that $H^2$ is roughly proportional to $\Phi^4$ 
but the mass squared is to $\Phi^2$. 
Hence, as the amplitude of $\Phi$ decays, the mass squared 
tends to be larger than $H^2$.\newline 

\noindent
\underline{\it\large {Detailed Analysis without Time Averaging}}$~$:$~$
We solve the above set of equations numerically 
varying the initial value of $\chi$. Then, we can evaluate 
the initial value dependence of $N_{H_r}$, the $e$-folding number 
until the total energy density reaches a certain specified value. 
However, the original model has the following shortcoming. 
As there are only two degrees of freedom corresponding 
to the homogeneous modes of $\phi$ and $\chi$-fields, 
the exchange of energy between them continues 
for ever. Then, occasionally the value of $\phi$ 
becomes exceptionally small. In this phase, the period between  
two neighboring zeros 
of $\mbox{\boldmath$\Phi$}\!\cdot\! \dot{\mbox{\boldmath$\Phi$}}$ 
becomes very long. If the specified target energy density 
is reached during this phase, $\Delta N$ becomes 
largely dependent on the choice of the final value of the Hubble rate
$H_r$. 
Then, interpretation of the results becomes difficult. 
In more realistic situations, the energy density 
carried by these long-wavelength modes will be redistributed to 
a large number of other degrees of freedom (shorter wavelength modes 
of these or other fields) sooner or later through rescattering.  

Here we assume two radiation fluids, $r_\phi$ and $r_\chi$. 
$\Phi^i$ decays into the component $r_i$ 
with decay constant $\Gamma_i$.  
The modified set of equations are given by 
\begin{eqnarray}
&& \ddot \Phi^i+(3H+\Gamma_i)\dot \Phi^i
    -{dV\over d\Phi^i}=0,\qquad
 \dot \rho_{r_i}=4H\rho_{r_i}+\Gamma_i (\Phi^i)^2, \qquad 
 H^2={\kappa^2\over 3}\left({1\over 2}
   \dot {\mbox{\boldmath$\Phi$}}^2+V+\sum_{i=1}^2\rho_{r_i}\right),\qquad 
\label{modifiedeq}
\end{eqnarray}
where $\rho_{r_i}$ is the energy density of the $i$-th component of 
radiation fluids. 

We consider the case with $g^2=2\lambda=0.02$ and 
$\Gamma_\phi=\Gamma_\chi=10^{-5}$. We set the 
initial condition at $\phi=1$ with $\dot\phi=\dot\chi=0$. 
Using the convention $\kappa^2/3=1$,
the initial Hubble rate is $H=0.05$, where 
we assume a negligible contribution to the initial energy density 
from the $\chi$-field. 
We integrate Eqs.~(\ref{modifiedeq}) 
until the Hubble rate goes down to $10^{-6}$, 
at which point the energy is almost completely transferred to 
the radiation fluids. In Fig.~1, we give a plot of the 
$e$-folding number $N$ between $H=0.05$ and $H=10^{-6}$ 
varying the initial value of $\chi$. We see 
small fluctuations of the $e$-folding number $N$ depending 
on the initial value of $\chi$. Each plot contains only 100 
sampling points with equal horizontal spacing. The amplitude of fluctuation 
is not $O(1)$ but still it is significant compared 
to that observed in the cosmic microwave background. 
An interesting and important point is that the dependence on 
the initial value of $\chi$ of this fluctuation is almost 
random. Two panels in Fig.~1 are plots with different 
range of the initial value of $\chi$. Although the scale of 
the horizontal axis is very different, these two plots 
look quite similar. 
\newline

\begin{minipage}{7cm}
\epsfxsize=6cm
\epsfbox{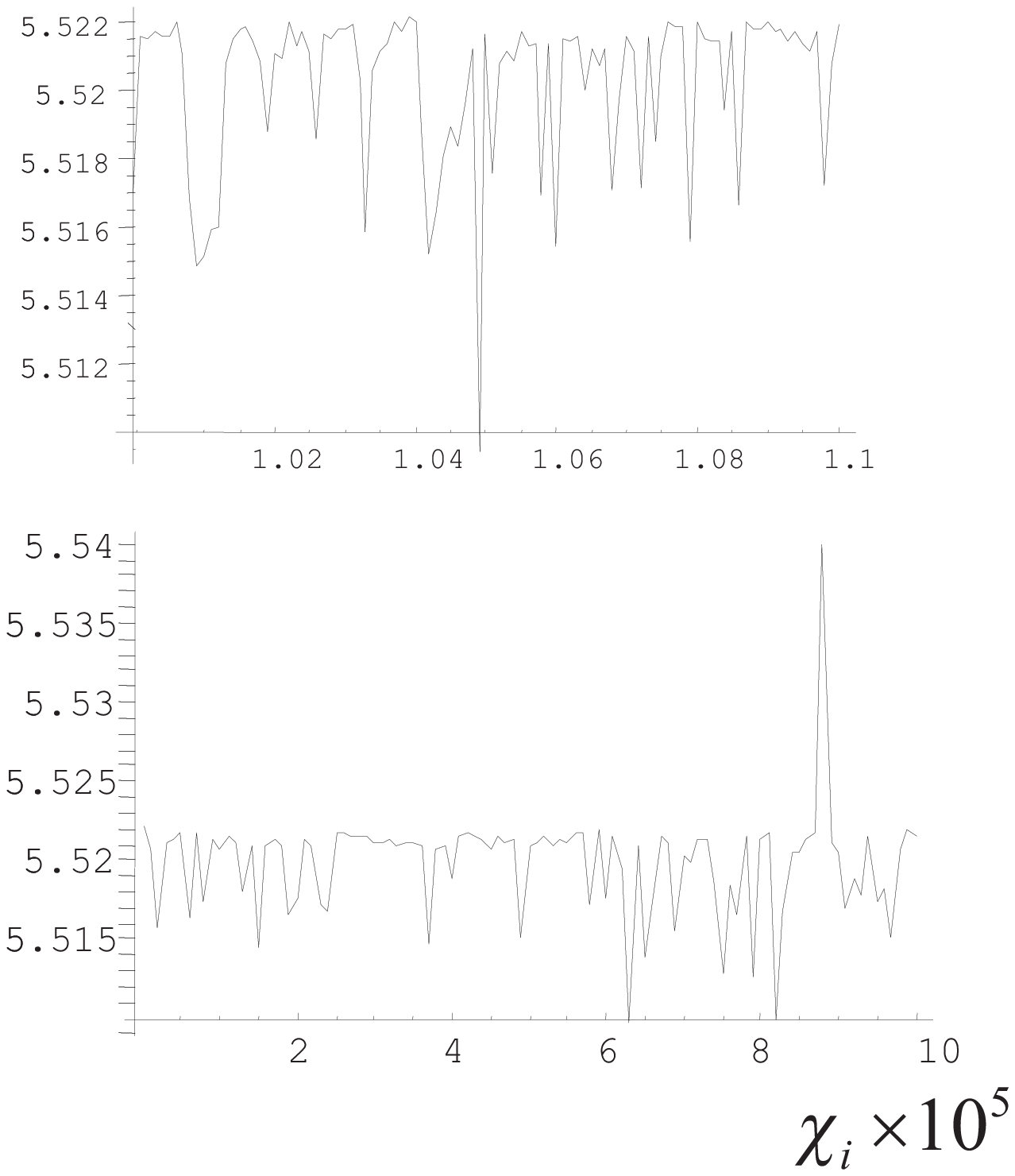}
\vspace{3mm}
{
\footnotesize 
\vbox{
\baselineskip=8pt 
Fig.1: Plots showing the initial value dependence of 
the $e$-folding number $N$. In each panel 100 discrete points 
are plotted. The scale of the horizontal axis is different, but 
these two plots look quite similar. This will mean that the distribution 
of $N$ is rather random.}}
\end{minipage}
\hspace{1cm}
\begin{minipage}{7cm}
\epsfxsize=5.5cm
\epsfbox{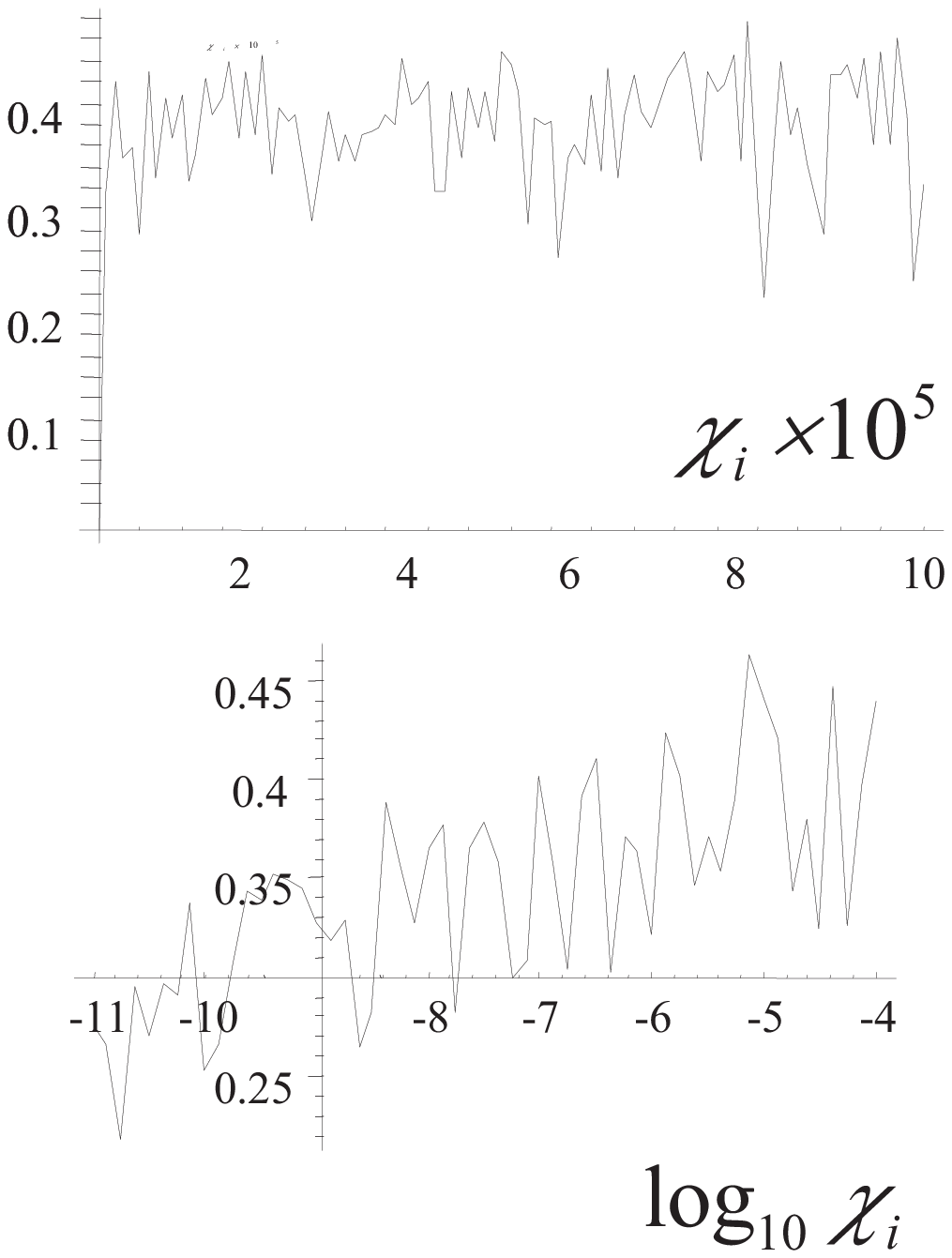}
{\footnotesize 
\vbox{
\baselineskip=8pt 
Fig.2: Similar plots to Fig.1 but the vertical axis is 
the fraction of the radiation energy
originating from the secondary field $\chi$. The panel (b) is 
a close-up view of the panel (a) for small initial value of $\chi$.
We can see a trend that the fraction increases as the initial 
value of $\chi$ increases.}}
\end{minipage}
\vspace{3mm}

The origin of these fluctuations can be understood as follows. 
If the oscillation of fields is fast compared to the change 
rate of $H$, the approximation using the time averaging 
must be good. The error will be exponentially 
suppressed. As mentioned above, however, the period of oscillation 
occasionally becomes very long if the motion of the scalar 
fields is trapped by chance in the region where the amplitude 
of oscillation in the direction of $\phi$-field is exceptionally 
small. In such situation, the approximation we made before is no 
longer valid. 
Since the motion of the two scalar fields are rather 
chaotic, the initially neighboring trajectories in the 
configuration space of $\phi$ and $\chi$ soon deviate. 
Thus, it is quite sensitive to the tiny difference in the 
initial condition whether this entrapment occurs or not. As a result 
the dependence on the initial condition is almost completely randomized. 

Then, the question is whether this amplification of fluctuations 
can be a process that generates significant fluctuations 
at the super-Hubble scale during preheating? 
Since the amplitude of the generated curvature perturbation is 
so sensitive to $\chi_i$, the variance 
seen in Fig.~1 will basically represent the variance 
of random fluctuations incoherent beyond the Hubble scale. 
If the fluctuations are completely random beyond a certain cutoff 
scale $k_c^{-1}$, 
we have $n=4$ spectrum for $k<k_c$ in general. 
(Here we use the notation that $n=1$ 
corresponds to the scale invariant spectrum.)

However, it is not certain whether a smoothing of $N(\chi)$, 
e.g., $\tilde N(\chi_i):=\int d\chi N(\chi) W(\chi-\chi_i)$ with 
an appropriate window function $W$ is tilted or not. 
Hence, the averaged value 
on very large scale may have an additional fluctuation 
other than the $n=4$ random component. The amplitude 
of this additional component is difficult to determine because 
we need to reduce the contamination of $n=4$ component. 
To do so, we need to compute the average over sufficiently large volume. 
This means that we need to calculate the 
value of $N$ for sufficiently large number of samples changing 
the initial value of $\chi$, which we have not done yet. 
Hence, we have not completely rejected the possibility 
of generating fluctuation coherent beyond the Hubble scale. 
However, the amplitude of this coherent fluctuation 
should be much more suppressed than the incoherent one. 

\vspace{3mm} 
\noindent
\underline{\it\large Generation of Curvature Perturbations}$~$:$~$

If the exact conformal invariance is broken by 
some reason, we can expect a curvature perturbation may be generated. 
The easiest way will be changing the decay process of the 
two fields. In the above calculation, we have introduced 
the decay terms by hand assuming that the both scalar fields 
decay into a radiation fluid with the same decay constant 
$\Gamma$. We distinguished two components of 
the radiation fluids depending on which scalar field becomes 
the origin. 
Suppose, 
one radiation fluid is composed of massive fields, 
although the mass is negligible during the preheating epoch. 
Then, sooner or later, 
the mass of this radiation fluid becomes larger than the 
Hubble rate, and the energy density of this field 
starts to behave like $\propto e^{-3N}$, while the other 
continues to evolve as $\propto e^{-4N}$. Therefore 
the isocurvature perturbation imprinted as the ratio 
$
 \displaystyle\xi:={\rho_{r_\chi}/( \rho_{r_\phi}+\rho_{r_\chi})}, 
$
is transformed into the curvature perturbation at this stage. 

Now let us look at the plot of this ratio as a function of 
the initial value of $\chi$. From Fig.~(2a), it seems that 
the dependence on $\chi_i$ is just random fluctuations. 
However, if we close up the portion for small $|\chi_i|$ (Fig.~(2b)), we 
can see clear dependence on $\chi_i$, which can be 
explained as follows. 
For small $\chi_i$, the decay to the radiation fluid 
becomes effective before the equi-partition between 
the homogeneous parts of the $\phi$ and $\chi$ fields 
is achieved by the parametric resonance. In this case, 
the decay to the radiation fluid is more efficiently 
through the $\phi$ field. As a result, the ratio $\xi$ 
becomes small for small $|\chi_i|$. 
This mechanism to generate large scale fluctuations 
is very efficient because the fluctuation continues to be 
amplified until the matter radiation equality. 
Hence, this process will be a possible mechanism 
to create primordial black holes.

\vspace{-3mm}
\section{Discussion}
\vspace{-2mm}
In this paper, applying the separate universe approach, 
we discussed generation of metric perturbations at super-Hubble 
scale during preheating. As an example, we considered the conformal 
two scalar model whose potential is given by Eq.~(\ref{conf-model}).  
Previously, it was argued that large amplitude of fluctuations
can be generated during the reheating process, but 
we found that the separate universe approach 
suggests relatively suppressed amplitude of fluctuations.  

We suspect that the origin of this discrepancy is in 
an overlooked point in previous analyses. 
Usually people decompose fields as 
(homogeneous background)+(perturbations), 
using the common homogeneous background for separated universes. 
Let's consider two largely-separated 
homogeneous patches, universe $A$ and universe $B$. 
Suppose that the preheating is efficient in the universe $A$ 
but inefficient in the universe $B$. 
Then, we physically expect that the back reaction to the 
inflaton oscillation is also efficient in the universe $A$.  
However, as long as we use the common homogeneous background, 
the back reaction works in the same way both in 
the universe $A$ and in the universe $B$. 
This means that artificial and 
acausal energy transfer from the universe $B$ to $A$ easily takes place 
in the treatment using the homogeneous background.

However, the above criticism to the previous works 
does not apply to the numerical studies. 
For example, in Ref.\cite{EasPar}, it was argued that super-Hubble modes are 
enhanced even in a single field model. 
They considered $\lambda\phi^4$ model. 
When they simulate the case that only a single mode 
in the resonance band is excited, there was no enhancement 
in super-Hubble modes. 
On the other hand, when all the modes in the same lowest 
resonance band are excited, there was enhancement in 
super-Hubble modes. 
Their explanation is as follows. The gauge invariant 
perturbations at super-Hubble scale are generated by 
the quadratic product of smaller scale perturbations 
in the resonance band. Namely, 
$
\zeta_{\bf k}\propto \int d^3k \chi({\bf k}')\chi({\bf k}-{\bf k}'). 
$
This integral becomes approximately constant for small $k$ 
less than the width of the resonance band $\Delta k$. 
For larger $k$, this integral is small because, when one of the 
arguments of $\chi$ is in the resonance band, the other is 
outside of it. 
So this enhancement mechanism does not apply for small scale modes. 
In this sense, the large scale modes are selectively enhanced. 

However, 
since $\zeta_{\bf k}\propto k^{-3/2}$ 
for the scale invariant spectrum, constant value of  
$\zeta_{\bf k}$ does not mean enhancement of larger wavelength mode. 
Therefore 
the mechanism of the enhancement of super-Hubble modes works 
only when the cutoff scale 
$a \Delta k^{-1}$ is much larger than the Hubble radius. 
We can easily see that this condition is not satisfied in this 
model. 
For the redefined field $\hat\phi=a\phi$, the field equation becomes 
$
\hat\phi{}''+\lambda\hat\phi{}^3-({a''/a})\hat\phi=0. 
$
Here we used the conformal time coordinate $\eta=\int a^{-1} dt$, and 
a prime denotes differentiation with respect to $\eta$. 
Since the radiation dominant universe is a good approximation 
for this model, we use $a=(t/t_0)^{1/2}$. Then, 
$
\eta=2t_0 a={1/ {\cal H}}={1/ a H}, 
$
and therefore $H=2t_0/\eta^2$. 
The amplitude of oscillation of $\hat\phi$ is constant. 
We can set the amplitude of $\lambda \hat\phi{}^2$ to unity without 
loss of generality. In fact, using the Friedmann equation, 
we have 
$
 \lambda\hat\phi{}^2 \approx {(8\pi G/ 3 a^2)}\lambda \phi_i^4
 ={\cal H}^2= 1/(2t_0 a)^2, 
$
where we have used the fact that the initial value of $\phi$,
$\phi_i$, is $O(m_{pl})$. Choosing $t_0$ appropriately, we 
can set $\eta_i=1$. This also means that the comoving wave number 
corresponding to the initial Hubble horizon ${\cal H}$ is 1.  

The resonance effect becomes important after $\mu\eta$ becomes $O(1)$, 
where $\mu$ is the growth rate of the amplitude due to parametric 
resonance. Hence, $\eta\approx \mu^{-1}$ when the parametric resonance 
becomes important. At this epoch, the comoving horizon scale is 
${\cal H}=\eta^{-1}\approx \mu$. Then, the condition
for the enhancement of the super-Hubble modes will 
be $\mu\gg \Delta k$. In this model $\mu<0.267$ and $\Delta k\approx
0.2$.  Therefore the condition is not satisfied. 

However, the plots in Ref.\cite{EasPar} show steep increase of 
power for small $k$. The expected flat spectrum, which does not 
mean scale invariance, is not seen. This is no contradiction. 
When they plot the figures, they plot only discrete numbers of $k$
modes and the plots do not have resolution less than $\Delta k$.  
This can be seen by looking at the shape of the resonance peak, 
which does not show the expected round shape. This means that 
the resolution is less than $\Delta k$. Since the mode $k\approx \Delta k$
is inside the Hubble scale after the resonance becomes efficient, 
actually only one mode outside the horizon scale is plotted in 
those figures. Therefore it is not strange that the expected 
plateau is absent in their plots.  
As mentioned before, 
this $n=4$ spectrum is common in the situation that 
there are random fluctuations of a large amplitude 
on small scales and seems to refute the claims of \cite{BV}. 
\newline

\baselineskip12pt
\end{document}